\documentclass[aps,prl,10pt,twocolumn]{revtex4-2}
\usepackage{amsmath}
\usepackage{amssymb}
\usepackage{adjustbox}
\usepackage{tikz}
\usetikzlibrary{quantikz2}

\DeclareFontFamily{U}{bbold}{}
\DeclareFontShape{U}{bbold}{m}{n}
 {
  <-5.5> s*[1.069] bbold5
  <5.5-6.5> s*[1.069] bbold6
  <6.5-7.5> s*[1.069] bbold7
  <7.5-8.5> s*[1.069] bbold8
  <8.5-9.5> s*[1.069] bbold9
  <9.5-11> s*[1.069] bbold10
  <11-15> s*[1.069] bbold12
  <15-> s*[1.069] bbold17
 }{}

\DeclareRobustCommand{\identity}{%
  \text{\usefont{U}{bbold}{m}{n}1}%
}

\usepackage{hyperref}
\usepackage[capitalise]{cleveref}
\usepackage[english]{babel}

\makeatletter
\def\bbl@set@language#1{%
  \edef\languagename{%
    \ifnum\escapechar=\expandafter`\string#1\@empty
    \else\string#1\@empty\fi}%
  \@ifundefined{babel@language@alias@\languagename}{}{%
    \edef\languagename{\@nameuse{babel@language@alias@\languagename}}%
  }%
  \select@language{\languagename}%
  \expandafter\ifx\csname date\languagename\endcsname\relax\else
    \if@filesw
      \protected@write\@auxout{}{\string\select@language{\languagename}}%
      \bbl@for\bbl@tempa\BabelContentsFiles{%
        \addtocontents{\bbl@tempa}{\xstring\select@language{\languagename}}}%
      \bbl@usehooks{write}{}%
    \fi
  \fi}
\newcommand{\DeclareLanguageAlias}[2]{%
  \global\@namedef{babel@language@alias@#1}{#2}%
}
\makeatother
\definecolor{color1bg}{HTML}{068DA9}
\definecolor{color2bg}{HTML}{7E1717}
\DeclareLanguageAlias{en}{english}

\begin{document}
\title{Efficient Post-Quantum Secured Blind Computation}
\author{Ethan Davies}
\affiliation{Information Security Group, Royal Holloway University of London, Egham, Surrey, TW20 0EX, UK.}
\email{ethan.davies.2021@live.rhul.ac.uk}
\author{Alastair Kay}
\affiliation{Department of Mathematics, Royal Holloway University of London, Egham, Surrey, TW20 0EX, UK.}
\email{alastair.kay@rhul.ac.uk}
\date{\today}
\begin{abstract}
In the medium term, quantum computing must tackle two key challenges: fault tolerance and security. Fault tolerance will be solved with sufficiently high quality experiments on large numbers of qubits, but the scale and complexity of these devices means that a cloud-based access model is likely to dominate. How can we risk evaluating valuable computations on an untrusted server? Here we detail a verifiable circuit-based model that only requires classical communication between parties. The server is blind to the details of the computation, which is computationally secure.
\end{abstract}
\maketitle

Cloud computing is now ubiquitous, and especially useful for quantum computers which, at least initially, will require specialist knowledge and equipment to run. If client Clio wants to employ the server Severus to perform her valuable quantum computation (a few minutes' calculation might contain a result that the world's largest supercomputers cannot perform in several years \cite{arute2019,zhong2020}), how can she trust that Severus won't just steal her answer, and how does she know that Severus has even performed the computation, given that verifying the answer may itself be a non-trivial challenge?

Blind quantum computation, in which Clio sends many single-qubit states to Severus, who uses them in a measurement-based protocol, provides Clio with information theoretic security over her input state, computation, and the outcome, only leaking to the server information about bounds on the size of the computation \cite{broadbent2009,fitzsimons2017}. This astounding result, which has also been implemented experimentally \cite{barz2012}, has spawned numerous attempts to remove the one element that inhibits wide-scale utilisation: the requirement on Clio to produce and send single qubits (known as being `semi-classical'). These have included, for example, playing two different servers against each other \cite{broadbent2009,reichardt2013,reichardt2012,natarajan2017,gheorghiu2017,gheorghiu2015,mckague2016,fitzsimons2015,ji2016}. One further downside of this approach is that Clio must build into her circuits all the complexity of a fault-tolerant computation, with the massive qubit overheads that this entails \cite{gidney2021}.

An alternative approach, requiring only a single server and \emph{classical} communication, necessitates the relaxation from information theoretic to computational security, relying on a cryptographic protocol that is believed secure against a resource-constrained quantum adversary. On the other hand, it can be built on top of a server's fault-tolerant quantum computer, of which Clio can be entirely oblivious. It has been established \cite{mahadev2018}, with improvements by \cite{brakerski2018}, that certain post-quantum cryptography schemes can be used to create a quantum fully homomorphic encryption (QFHE) scheme which produces an encrypted computation on the server. While these papers proved existence and security, \emph{how} such a quantum computation proceeds was not detailed. In this paper, we make these steps explicit. We then extend the central construction of \cite{mahadev2018} from an encrypted controlled-\textsc{not} to an encrypted universal gate set, permitting a more direct, resource efficient, implementation. Post-quantum cryptography has also been used to directly replicate measurement-based blind computation \cite{cojocaru2021,cojocaru2019,test}. Our method is also more resource efficient than these.

\textbf{Overview of QFHE-based Blindness}: Conceptually, \cite{mahadev2018,brakerski2018} demand a \emph{fixed} computational scheme built entirely out of Clifford gates and Toffolis; the server's circuit only depends on the size of the client's computation, with the computation being entirely determined by the initial state, i.e.\ this will proceed like a quantum cellular automaton (QCA) \cite{shepherd2006}. Clio provides three things: A bit string specifying the initial (basis) state of the computation which has a one-time pad attached to it, $Z^zX^x$ (Here, $X^x$ conveys the application of Pauli $X$ to qubits $i$ where the bit string $x\in\{0,1\}^n$ satisfies $x_i=1$.), an encryption of the one-time pad (padding), $\text{Enc}(z,x)$ and an evaluation key for her encryption scheme. Severus initialises his qubits in the specified basis state (including the one-time pad), and starts to apply the circuit. Every time a Clifford gate $U$ is applied, the pad must be updated to $UZ^zX^xU^\dagger=Z^{z'}X^{x'}$ \cite{childs2005}. With the evaluation key, homomorphic encryption schemes allow this calculation directly on the encrypted data $\text{Enc}(z,x)\mapsto \text{Enc}(z',x')$. Every time a Toffoli $\Lambda$ is applied, we have
$$
\Lambda Z^zX^x\ket{\psi}=V_{z',x'}Z^{z'}X^{x'}(\Lambda\ket{\psi}).
$$
Not only does the padding update, with Severus computing $\text{Enc}(z,x)\mapsto \text{Enc}(z',x')$, but an additional $V^\dagger$ must be applied to the quantum register. In this case, $V$ comprises some corrective controlled-\textsc{not}s determined by the padding. The key tool in \cite{mahadev2018} is an \emph{encrypted} controlled-\textsc{not} allowing Severus to apply the required corrections based only on the encryption of the pad. After all elements of the circuit have been applied, Severus measures every qubit in the standard basis. This is the padded answer, which is sent to Clio who, knowing the pad, can extract the answer to her computation. Note that, for a general gate $\Lambda$ at level $k$ in the Clifford hierarchy \cite{gottesman1999}, the correction $V^\dagger$ is at level $k-1$ in the hierarchy.

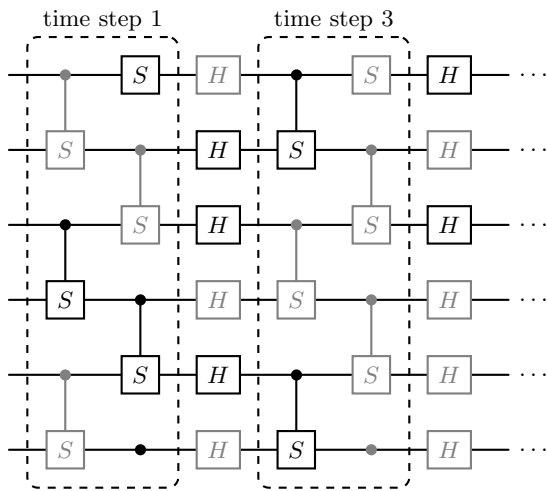
\begin{figure}
\centering
\begin{quantikz}
& \ctrl[style={draw=gray,fill=gray}]{1}\gategroup[6,steps=2,style={dashed, rounded corners}]{time step 1} & \gate{S} & \gate[style={draw=gray},label style={gray}]{H} & \ctrl{1}\gategroup[6,steps=2,style={dashed, rounded corners}]{time step 3} & \gate[style={draw=gray},label style={gray}]{S} & \gate{H} & \rstick{\ldots} \\ 
& \gate[style={draw=gray},label style={gray}]{S} & \ctrl[style={draw=gray,fill=gray}]{1} & \gate{H} & \gate{S} & \ctrl[style={draw=gray,fill=gray}]{1} & \gate[style={draw=gray},label style={gray}]{H} & \rstick{\ldots} \\
& \ctrl{1} & \gate[style={draw=gray},label style={gray}]{S} & \gate{H} & \ctrl[style={draw=gray,fill=gray}]{1} & \gate[style={draw=gray},label style={gray}]{S} & \gate{H} & \rstick{\ldots} \\ 
& \gate{S} & \ctrl{1} & \gate[style={draw=gray},label style={gray}]{H} & \gate[style={draw=gray},label style={gray}]{S} & \ctrl[style={draw=gray,fill=gray}]{1} & \gate[style={draw=gray},label style={gray}]{H} & \rstick{\ldots} \\
& \ctrl[style={draw=gray,fill=gray}]{1} & \gate{S} & \gate{H} & \ctrl{1} & \gate[style={draw=gray},label style={gray}]{S} & \gate[style={draw=gray},label style={gray}]{H} & \rstick{\ldots} \\ 
& \gate[style={draw=gray},label style={gray}]{S} & \control{} & \gate[style={draw=gray},label style={gray}]{H} & \gate{S} & \control[style={draw=gray,fill=gray}]{} & \gate[style={draw=gray},label style={gray}]{H} & \rstick{\ldots}
\end{quantikz}
\caption{Subset of qubits in a larger array with repeated applications of controlled-$S$ and Hadamard. Relative to this repeating pattern, the gates Severus actually applies (black) at time step $l$ are specified by a bit string $y_l\oplus x_l$, a combination of the ideal gates $y_l$ and a one-time pad $x_l$.}\label{fig:basic_circuit}
\vspace{-0.5cm}
\end{figure}

\begin{figure}
(a)\begin{quantikz}[wire types={q,q,q,c,q},align equals at=1,column sep=0.3cm,row sep=0.3cm]
& \ctrl{1} &&&&\ctrl{1} & \\
& \ctrl{1} &&&&\ctrl{1} & \\
\lstick{\ket{0}}& \targ{} & \ctrl{1}&&\ctrl{1} & \targ{} & \\
\lstick{c} &&\ctrl{1} && \ctrl{1} && \\
\lstick{\ket{0}}&&\targ{}&\gate{S}&\targ{}&&
\end{quantikz}

(b)
\begin{quantikz}[wire types={q,c,q},align equals at=1,column sep=0.3cm,row sep=0.3cm]
& \targ{} & \ctrl{1} &\targ{} &&\targ{}&\ctrl{1}&\targ{}&\\
\lstick{c}&&\ctrl{1}&& &&\ctrl{1}&& \\
\lstick{\ket{0}}&\ctrl{-2}& \targ{} &\ctrl{-2}& \gate{H} & \ctrl{-2} & \targ{} & \ctrl{-2}&
\end{quantikz}
\caption{Creation of (a) controlled-$S$ and (b) Hadamard both controlled off qubit $c$ prepared in a basis state, up to global phases, using Toffoli and Cliffords. Any gate controlled off $c$ can be replaced by a classical decision to apply an uncontrolled version provided we still propagate Paulis and apply encrypted c-\textsc{not}s as if the substitution hadn't been made.}\label{fig:gate_sim}
\vspace{-0.5cm}
\end{figure}
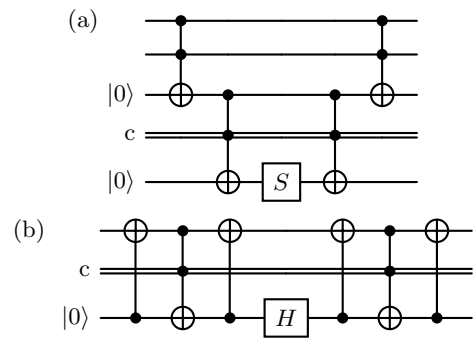

\textbf{Explicit Computational Scheme}: In \cite{mahadev2018} it was proved that this methodology results in a universal quantum computation if the circuits are designed appropriately. No such designs were provided. We now detail an explicit method using the universal gate set comprising Hadamard and controlled-$S$ (acting only between nearest neighbours) \cite{kitaev1997}. Consider a computational register of $n$ qubits which has $m$ alternating layers of controlled-$S$ and $H$ applied across all qubits, as depicted in \cref{fig:basic_circuit}. If we could choose which of these gates to apply, or not, we could implement our desired circuit. Let $y_l\in\{0,1\}^{n}$ specify which gates in layer $l$ should be on/off. The initial state of the computation that Clio supplies to Severus is $(0^{n}y_1y_2\ldots y_m)\oplus x$, where $x$ is the one-time pad. Severus holds two quantum registers, the first storing the $n$ computational qubits, and the second holding all the $mn$ padded steps $y_l$. On a single layer/step, controlled-controlled-$S$ (controlled-$H$) is implemented between a pair of neighbouring (a single) computational qubits controlled off one program qubit from the current layer. We can build both of these gates out of Clifford + Toffoli with the help of two (one) ancillas, as seen in \cref{fig:gate_sim}.



This na\"ive version creates a QCA \cite{shepherd2006} where the entire computation is specified in a very large quantum state. Of course, the program data is entirely classical. Instead of holding that data in a quantum register, Severus should just choose to apply the gate based on the classical data he holds. However, care must be taken to not miss important corrections due to the padding. If his bit value $c$ suggests that controlled-$S$ (for example) need not be applied, the entire sequence must still be applied as new Pauli corrections and encrypted c-\textsc{not}s can arise at each step, meaning that some gates do not cancel in the way the circuit diagram might suggest.

Moving the program bits from a quantum to a classical register massively reduces the overhead in the number of qubits (now we only need $O(n)$, not $O(nm)$) although it has little impact on the main cost: 3 encrypted controlled-\textsc{not}s applied after every Toffoli. A further inefficiency is that the gate set ($H$, c-$S$) is non-standard, and has to be constructed out of Toffoli + Clifford which are not particularly natural in a fault-tolerance scenario: you would not include the 3-qubit Toffoli in a fault-tolerant gate set because it has a massive impact on the threshold. Instead, one would have to synthesise this from the available gates, adding to the overhead. 

The gadget in \cref{fig:gate_sim} can replace $H$ with any other Clifford gate to achieve a blind version of that gate. Gates in higher levels of the Clifford hierarchy may also be used but this requires lower level corrections which can be implemented using the same set-up until all required corrections are dealt with.

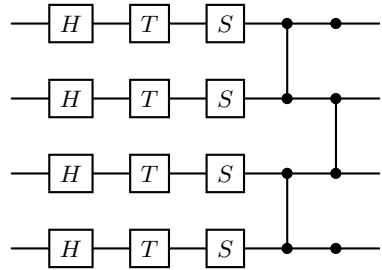
\begin{figure}
\centering
\begin{quantikz}
& \gate{H} & \gate{T} & \gate{S} & \ctrl{1} &\control{}& \\
& \gate{H} & \gate{T} & \gate{S} & \control{} &\ctrl{1}& \\
& \gate{H} & \gate{T} & \gate{S} & \ctrl{1} &\control{}& \\
& \gate{H} & \gate{T} & \gate{S} & \control{} &\control{}&
\end{quantikz}
\caption{One layer, which would be repeated many times to achieve a computation. Each gate is individually switched on/off by an encrypted bit. Each gate is immediately followed by the gate that offers any required non-Pauli correction.}\label{fig:newlayer}
\vspace{-0.5cm}
\end{figure}

\textbf{Direct Computational Scheme}: We propose replacing the transmission of the one-time padded computation and the encrypted pad with simply an encryption of each computational layer, $y_l$ as independent one-bit encodings. These are used to determine the application of an array of encrypted unitaries directly. The initial state may still be padded, and Clio keeps track of this padding as it propagates through the various gates, which add their own Pauli corrections without ever revealing any part to Severus. We envisage, for instance, a computation that proceeds by many repetitions of the circuit in \cref{fig:newlayer}, with every gate being independently turned on or off. All gates in this sequence are Clifford, except for $T$, although there is great flexibility in potential repeating patterns for tuning to specific algorithmic or hardware requirements. Since $TZT^\dagger=Z$ and $TXT^\dagger=S^\dagger ZX$ (up to an irrelevant global phase) the only non-Pauli correction is an $S$ which we incorporate in the next step. While any single-qubit Clifford unitary can be implemented in an encrypted manner using two encrypted c-\textsc{not}s via \cref{fig:gate_sim}, we will give direct constructions in the following section for each unitary in \cref{fig:newlayer}, only requiring the same complexity as one encrypted c-\textsc{not}.



\begin{figure}
\centering
\begin{quantikz}[wire types={q,q,q,b,b},classical gap=0.05cm,column sep=0.3cm,row sep=0.3cm]
\lstick{$q_1$} & & \gate{V}&& &&&[-0.3cm]\\
\lstick{$q_2$}&&&\gate[4][1cm]{U_f}\gateinput{$a$} &&&& \\
\lstick{\ket{0}}\gategroup[steps=3,style={fill=color1bg!50,draw=none,rounded corners,xshift=-0.3cm},background]{} & \gate{H}&\ctrl{-2} &\gateinput{$d$}&\gate{H} & \meter{}&\rstick[2]{$R$}\setwiretype{n} \\
\lstick{$\ket{0}^{\otimes n}$} & \gate{H}& &\gateinput{$r$}& \gate{H} &\meter{}&\setwiretype{n} \\
\lstick{$\ket{0}^{\otimes m}$} &&&\gateinput{$y$}& \meter{}\ar[r]& \rstick{$Y$}\setwiretype{n}
\end{quantikz}
\caption{Implementation of the encrypted c-$V$, controlled by $q_2$, targeting $q_1$, and enacted depending on the value $\hat d$ which is never revealed to the server. If $V$ is Pauli, no corrections are required beyond updates to the Pauli padding. When $V=e^{i\theta}\identity$, the shaded region can be replaced by magic state preparation.}\label{fig:enc_cnot}
\vspace{-0.5cm}
\end{figure}
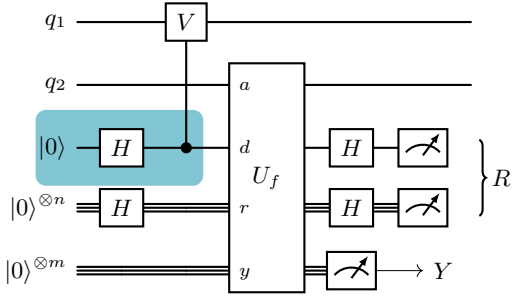

\textbf{Encrypted Unitaries}: Using the post-quantum public key cryptography Learning With Errors (LWE) \cite{regev2009}, Clio samples a public key \textsf{pk} and a secret key \textsf{sk} with associated security parameter $\lambda$. The public key is used to encrypt a bit $d$ using some randomness $r$, $\text{Enc}_{\textsf{pk}}(d,r)$. We make the computational assumption that this can only be decrypted in reasonable time with access to the secret key \textsf{sk}. The encryption algorithm further satisfies:
\begin{equation}\label{eq:key}
\text{Enc}_{\textsf{pk}}(d,r)+\text{Enc}_{\textsf{pk}}(d',r')=\text{Enc}_{\textsf{pk}}(d\oplus d',r''),
\end{equation}
where the ``+'' operation can be performed with knowledge of the public key. Let $\hat d$ be a bit held by Clio, determining the action of a single gate. She sends her encryption of it, $\hat y=\text{Enc}_{\textsf{pk}}(\hat d,r)$, to Severus along with the public key. He defines two injective functions 
$$
f_0(d,r)=\text{Enc}_{\textsf{pk}}(d,r),\qquad f_1(d,r)=\text{Enc}_{\textsf{pk}}(d,r)+\hat y,
$$
and builds them into a unitary $U_f$ with the action
$$
U_f\ket{a}\ket{d}\ket{r}\ket{y}=\ket{a}\ket{d}\ket{r}\ket{y+f_a(d,r)}.
$$
We use this in the circuit of \cref{fig:enc_cnot} to implement a controlled-$V$ between two qubits determined by the value $\hat d$. Note that after the measurement on the final register, we have collapsed the other registers onto values such that $Y=f_0(d_0,r_0)=f_1(d_1,r_1)$ \footnote{The procedure can fail at this point if the two different $r$ values are different lengths, one being beyond the size of the $r$ register. Under the assumption that Yao’s XOR
Lemma also applies for one-round protocols (with classical messages) against a quantum adversary, the probability of failure is negligibly small \cite{cojocaru2019}. We can thus simply restart the entire computation if it fails, without affecting the circuit complexity.}. By \cref{eq:key}, we know that $f_1(d_1,r_1)=f_0(d_1\oplus\hat d,r')$. Injectivity thus implies that $d_0=d_1\oplus\hat d$, which is conveniently rewritten as $d_i=d_0\oplus(i\times \hat d)$. By this token, the final action is
\begin{equation}\label{eq:cVcorrections}
Z_2^{R\cdot(\hat d\|r_0\oplus r_1)}({c_2-V_1})^{-2d_0\hat d}V_1^{d_0}(c_2-V_1)^{\hat d}.
\end{equation}
While Severus does not know $\hat d,d_0,r_0,r_1$, Clio can apply the decryption (trapdoor) function on the value $Y$ to determine them. If c-$V$  is in the Clifford hierarchy, then the hierarchy of corrections is finite, the c-$V^2$ terms can be dealt with similarly and the remaining corrections will also be in the Clifford hierarchy and can be dealt with as in the explicit computational scheme. If $V$ is a Pauli operator, the only further corrections required are an adaptation to the padding, giving a particularly convenient encrypted c-\textsc{not}, c-$Z$.

\begin{figure}
\centering
\begin{quantikz}[wire types={q,q,q,b,b},classical gap=0.05cm,column sep=0.3cm,row sep=0.3cm]
\lstick{\ket{\psi}} & &&  &\gate{iY}&&&[-0.3cm]\\
\lstick{\ket{+}}  &         &\gate[4][1cm]{U_f}\gateinput{$a$}&\targ{}  & \ctrl{-1}&\gate{H}&\meter{}\ar[rr]&\setwiretype{n}&\rstick{$A$} \\
\lstick{\ket{0}}  & \gate{H}&\gateinput{$d$}                  &\ctrl{-1}&          &\gate{H}& \meter{}      &\rstick[2]{$R$}\setwiretype{n} \\
\lstick{$\ket{0}^{\otimes n}$} & \gate{H} &\gateinput{$r$}&&& \gate{H} &\meter{}&\setwiretype{n} \\
\lstick{$\ket{0}^{\otimes m}$} &&\gateinput{$y$}& \meter{}\ar[r]& \rstick{$Y$}\setwiretype{n}
\end{quantikz}
\caption{Implementation of the encrypted Hadamard, enacted depending on the value $\hat d$.}\label{fig:enc_had}
\vspace{-0.5cm}
\end{figure}
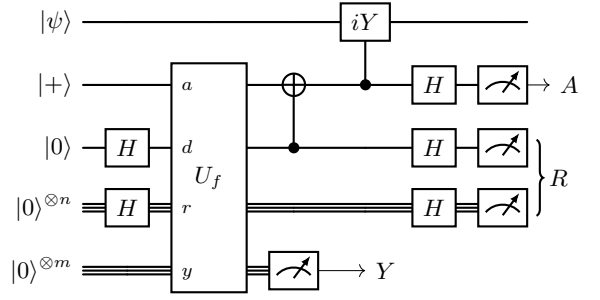

Encrypted phase gates derive from the same construction: $V=e^{i\theta}\identity$ is equivalent to a phase gate $P$ applied to the control only. This has the added benefit that the input state on $d$ is just a magic state for that phase gate. By \cref{eq:cVcorrections}, the action is
$
Z^{R\cdot(\hat d\|r_0\oplus r_1)}P^{-2d_0\hat d}P^{\hat d}.
$
As well as having an encrypted $P$, this also provides the required correction for encrypted controlled-$P$.
The universal set of encrypted gates for \cref{fig:newlayer} is completed by the Hadamard construction given in \cref{fig:enc_had}, which acts as
$$
Z^{\bar{\hat d}}Y^{(\hat d\|r_0\oplus r_1)\cdot R\oplus A\oplus (d_0\oplus A)\hat d}H^{\bar{\hat d}}.
$$
Strictly, the encrypted Hadamard is not necessary. One could proceed to apply Hadamard to every qubit at each required step. Much as for the brickwork state \cite{broadbent2009}, three rounds provide the ability to switch on/off a Hadamard on each qubit by suitable choices of the $S$ being on/off. In the context of gate synthesis, the cost is negligible.

\textbf{Security}:
We prove our scheme is secure in that if there exists a quantum polynomial time (QPT) algorithm $d' \leftarrow \mathcal{A}(\textsf{pk},y_i)
$ that can distinguish between any two possible quantum circuits with probability better than $1/2+1/\text{poly}(\lambda)$, then there also exists a QPT algorithm $d \leftarrow \mathcal{B}(\text{Enc}(\hat{d}))$ such that $P(d=\hat{d}) >1/2+1/\text{poly}(\lambda)$, i.e.\ we would be able to break the encryption of the public key. This is impossible by assumption.

\textbf{Proof}:
To construct an algorithm for $\mathcal{B}$, we first deal with the case where all blind gates need no further corrections (i.e.\ Clifford only), and then show how the proof can be adapted to allow for the necessary corrections.

\begin{enumerate}
\item $\mathcal{B}$ receives as input $\textsf{pk}$ and $\hat{y}=\text{Enc}_{\textsf{pk}}(\hat{d},r)$. His challenge is to determine $\hat d$.
\item $\mathcal{B}$ chooses any two circuits $C_0, C_1$ of the same size implementable by $\mathcal{A}$.
\item For each gate $i$ in $C_0$, $\mathcal{B}$ produces encryptions $y_i = \text{Enc}_{\textsf{pk}} ({C_0}_i)$.
\item $\mathcal{B}$ then computes:
\begin{align*}
\hat{y}_i &= y_i + ({C_0}_i\oplus {C_1}_i)\hat{y}\\
&= \text{Enc}_{\textsf{pk}}({C_0}_i)+ ({C_0}_i\oplus {C_1}_i)\text{Enc}_{\textsf{pk}}(\hat{d})\\
&= \text{Enc}_{\textsf{pk}}({C_0}_i \oplus \hat{d}\times ({C_0}_i\oplus {C_1}_i))\\
&= \text{Enc}_{\textsf{pk}}({C_{\hat{d}}}_i).
\end{align*}
\item $\mathcal{B}$ runs $\mathcal{A}$ and computes $d' \leftarrow \mathcal{A}(\textsf{pk},\hat{y}_i)$, i.e.\ outputs on $d'$ which of the two circuits it was.
\item $\mathcal{B}$ then outputs $d'$.
\end{enumerate} 

$\mathcal{B}$ succeeds when $\mathcal{A}$ succeeds, so our cryptographic assumption that $\mathcal{B}$ cannot succeed in determining $\hat d$ implies that $\mathcal{A}$ cannot succeed, completing our simplified proof. So far, these circuits only consist of Clifford gates and cannot produce any quantum advantage \cite{aaronson2004}. Unfortunately, $\mathcal{B}$ has no way to compute the required corrections for non-Clifford gates as that would involve access to the trapdoor (decryption) function. We circumvent this by having $\mathcal{B}$ generate a random encryption and use that in its place. Since the public key encryption scheme has semantic security, replacing the correct encryption with a randomly generated one can only change the probability distribution by a negligible amount. Since there are only $\text{poly}(\lambda)$ corrections needed, the success for $\mathcal{B}$ can be kept negligibly close to the case in which $\mathcal{B}$ succeeds with the correct encryptions for the needed corrections.

\textbf{Verification}: Verification of a quantum computation is also an important task -- as much as the server claims to be performing a quantum computation, how sure can we be of this fact? We compute because we don't know the answer, and the answer may not be easily verifiable. The tricks are well-known in blind computation: build in trap wires. In various places throughout the computation, create states that should have specific values. These could be single-qubit states or perhaps multi-qubit GHZ states, which provide an all-versus-nothing proof that the server is properly producing entangled states \cite{greenberger1989}, and must therefore be performing faithfully as, by the blindness properties, they cannot know where these traps are.

\textbf{Conclusions}: We have explicitly specified a computational scheme based on \cite{mahadev2018} that allows universal quantum computation via QFHE. The work of \cite{mahadev2018} was subsequently improved in \cite{brakerski2018} to use standard, well-studied, security assumptions of the LWE problem. For pedagogical purposes, we have presented our results with respect to the original work. In the Appendix, we show equivalent constructions in the improved case of \cite{brakerski2018}.

By relaxing the QFHE requirements (incorporating the feedback of Severus' measurement results) and introducing new encrypted unitary gadgets, we have provided a more direct method for blind computation with lower overhead: the comparison can be made directly by using the layer structure of \cref{fig:basic_circuit}. We have to add extra corrective controlled-phase and $S$ layers. Nevertheless, for each repetition of the structure, we need to apply $U_f$ four times per qubit, and with the benefit of including additional gates, thereby reducing the required circuit depth. The indirect method requires 18 applications of $U_f$.

\begin{figure}
\adjustbox{scale=0.6}{
\begin{tikzpicture}[client/.style={draw=black,rounded rectangle, fill=color1bg!50,minimum width=5.8cm,minimum height=0.8cm},server/.style={draw=black,rounded rectangle, fill=color2bg!50,minimum width=5.8cm,minimum height=0.8cm}]
\draw (-3,0.5) node {(b)};
\draw (0,0) node [client] {High-level programming language};
\draw (0,-1) node [client] {Gate model};
\draw (0,-2) node (client) [client] {Client's security \& verification};
\draw (0,-4.5) node (server) [server] {Fault-tolerance};
\draw (0,-5.5) node [server] {Physical level: Pulse sequences};
\draw [arrows = {-Stealth[harpoon]},thick] ($(client.south)+(0.05,-0.1)$) -- ($(server.north)+(0.05,0.1)$);
\draw [arrows = {-Stealth[harpoon]},thick] ($(server.north)+(-0.05,0.1)$) -- ($(client.south)+(-0.05,-0.1)$) node [midway, anchor=east,align=right] {Classical communication\\Computational security};
\begin{scope}[xshift=-8cm]
\draw (-3,0.5) node {(a)};
\draw (0,0) node [client] {High-level programming language};
\draw (0,-1) node [client] {Gate model};
\draw (0,-3) node (client) [client] {Client's security \& verification};
\draw (0,-2) node [server] {Fault-tolerance};
\draw (0,-5.5) node (server) [server] {Physical level: Pulse sequences};
\draw [arrows = {-Stealth[harpoon]},thick] ($(client.south)+(0.05,-0.1)$) -- ($(server.north)+(0.05,0.1)$) node [midway, anchor=west,align=left] {Quantum communication\\Information Theoretic security};
\draw [arrows = {-Stealth[harpoon]},thick] ($(server.north)+(-0.05,0.1)$) -- ($(client.south)+(-0.05,-0.1)$);
\end{scope}
\end{tikzpicture}
}
\caption{Computational stack for (a) measurement-based blind computation and (b) current proposal. In the current architecture, the server does not have to reveal anything about its capabilities to the client, who does not need to worry about the complexities of implementing fault tolerance.}\label{fig:architecture}
\vspace{-0.5cm}
\end{figure}
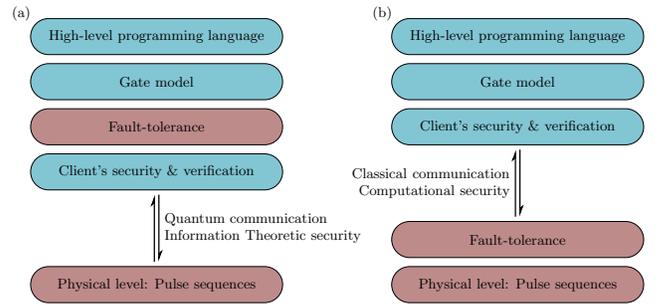

Direct recreation of measurement-based blind computation using post-quantum cryptography and computational security \cite{cojocaru2021,cojocaru2019,test} requires two secure bits per qubit to be created, and hence a total of 16 to produce the brickwork state \cite{fitzsimons2017,broadbent2009}. We can reproduce the same structure with just 13 encrypted bits, essentially due to the greater efficiency of the two-qubit gate, and with the additional freedom to independently choose the single qubit rotations and the two-qubit gate in the same unit.

Our preferred circuit design in \cref{fig:newlayer} utilises the more standard (finite) gate set of $T$, $H$ and a c-$Z$. In fact, since any phase gate $P$ satisfies $PZP^\dagger=Z$ and $PXP^\dagger=P^2X$, we could implement arbitrary chains of phase angles $\frac{\pi}{2^k}$, allowing for much shorter depth gate synthesis. However, since we envisage this sitting on top of a fault-tolerant implementation which will have a specific finite gate set (e.g.\ $T,S,H,\text{c-}\textsc{not},\text{c-}Z$) it will ultimately be most efficient to construct the layered structure out of these specific gates, which was our motivation in choosing \cref{fig:newlayer}. Indeed, this structure is well tuned to the task of synthesis: with the controlled-phase gates removed, the single-qubit gates naturally describe a sequence of $HT^{k_1}HT^{k_2}HT^{k_2}\ldots$ for $k_i\in\{0,1,2,3\}$, which is exactly the structure required to synthesise single qubit unitaries from $H$ and $T$ with maximum efficiency \cite{kliuchnikov2012,kliuchnikov2015}. Moreover, option to use magic states to implement the phase gate is built in.

One of the major benefits of reducing the semi-classical client to a classical one is depicted in \cref{fig:architecture}. In our scheme, the client only needs to specify a computation that sits on top of a fully fault-tolerant quantum computer. She can assume that the computation proceeds perfectly and be entirely oblivious to the error correction procedures implemented by Severus. On the other hand, a semi-classical Clio would be supplying Severus with single-qubit states. These cannot be protected in an error correcting code, and consequently are inherently noisy. Even if they could be supplied to the server with a high enough quality to surpass some critical threshold, the client would have to build appropriate error correction of those states into the structure of their measurement-based computation \cite{raussendorf2007}, adding significantly to the complexity for the client and to the resource overheads on the server. Most likely, the client's circuits would operate at the level of physical qubits on the server, building in the full complexity of fault tolerance and requiring knowledge of the server's noise properties. Our method shows that Clio can be relieved of this immense burden.

\bibliography{cs_blind_Ethan_edits_shorter_abstractNotes.bib}
\appendix
\section{Gate Constructions with Improved Security}\label{sec:appendix}

Here we discuss the more general framework for how the blind c-\textsc{not}$^{\hat{d}}$ works \cite{brakerski2018} and show that it too can be generalised easily to a $\text{c-}V^{\hat{d}}$ gate. This construction with the correct assumptions will allow for the gate to occur blindly, with clear indication where the cryptography plays a role.

The more general blind gate requires a family of functions $\mathcal{F}=\{f_{k,d}:\mathcal{X}\rightarrow \mathcal{Y}\}_{k \in K, d\in\{0,1\}}$ with security parameter $\lambda$ and the following properties:

\begin{itemize}
\item For all QPT algorithms $\mathcal{A}$, $$P(\hat{d} = d |d \leftarrow \mathcal{A}\,(f_{k,\hat{d}}), f_{k,\hat{d}} \leftarrow \mathcal{F})= \frac{1}{2} + \frac{1}{\text{negl}(\lambda)}.$$
\item there exist states $\ket{\phi_i}$ for $i\in\{0,1\}$ which are superpositions of the basis states $\mathcal{X}$ and can be made via $U_i\ket{0}^{\otimes n}$. It is then natural to reversibly compute $f_{k,\hat{d}}$ on $\ket{\phi_i}$ (using $U_f$, see \cref{fig:gen}) and measure the additional register. This collapses us onto one of the family of (unnormalised) states $\ket{\phi_{i,y}}=\mathcal{P}_{f_{k,\hat{d}}(\cdot)=y}\ket{\phi_i}$. In order to make progress with these states, additional global properties must hold for all $y$:
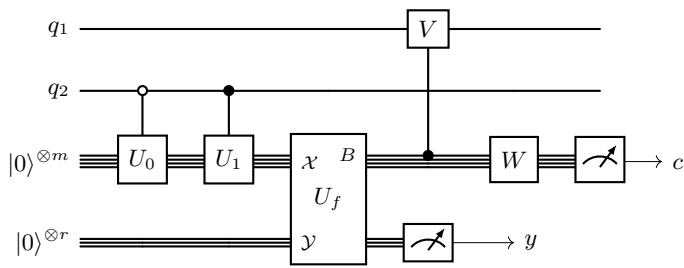
\begin{figure}[!tb]
\centering
\begin{quantikz}[wire types={q,q,q,b,b},classical gap=0.05cm]
\lstick{$q_1$}& &  & & \gate{V} & & \\
\lstick{$q_2$} & \octrl{1} & \ctrl{1}& & & &\\
& \gate[2]{U_0} &\gate[2]{U_1} & \gate[3][1cm]{U_f}\gateoutput{$B$} & \ctrl{-2} & \gate[2]{W} & \meter[2]{} \\[-0.95cm]
\lstick{$\ket{0}^{\otimes m}$} &&&\gateinput{$\mathcal{X}$}&&&  \ar[r,shift left=0.025cm]& \rstick{$c$}\setwiretype{n}\\
\lstick{$\ket{0}^{\otimes r}$}& & &\gateinput{$\mathcal{Y}$} &  \meter{}\ar[r]&\rstick{$y$} \setwiretype{n}
\end{quantikz}
\caption{Generalised setting to perform an encrypted gate. The c-$V$ gate is controlled off bit $B$ in the $m$-qubit register.}\label{fig:gen}
\end{figure}
\begin{enumerate}
\item There exists a single bit $B$ in $\ket{\phi_{i,y}}$ which is $\ket{b_y \oplus i \times \hat{d}}$. Moreover given $y$ and the trapdoor for $f$, $b_y$ can be computed efficiently.
\item $\exists W\, \forall y,i \, \exists z_i$ satisfying $\ket{\phi_{i,y}}=W^\dagger Z^{z_i} W\ket{\phi_{0,y}}$ such that, with access to $y$ and the trapdoor, $z_i$ can computed efficiently. Note that this imposes that $\ket{\phi_{i,y}}$ have the same normalisation for all $i$.
\end{enumerate}
\end{itemize}
With these ingredients we can enact an encrypted $c-V^{\hat{d}}$, using the circuit shown in \cref{fig:gen} with the final action on the state being 
$$
Z_2^{c\cdot(z_0\oplus z_1)}({c_2-V_1})^{-2b_y\hat d}V_1^{b_y}(c_2-V_1)^{\hat d},
$$
as required. Brakerski \cite{brakerski2018} was able to demonstrate that a different encryption scheme to Mahadev \cite{mahadev2018} could be used such that these properties hold and is believed to be more secure for the same security parameter.

\end{document}